\documentclass[draftclsnofoot,journal]{IEEEtran}
\usepackage[]{graphicx}
\begin{document}

\title{Spin State Read-Out by Quantum Jump Technique: for the Purpose of Quantum Computing}

\author{E.~Pazy \IEEEmembership{Department of Physics, Ben-Gurion University of the Negev,
Beer-Sheva 84105, Israel}\\
        T.~Calarco 
        and~P.~Zoller \IEEEmembership{Institut f\"ur Theoretische Physik, Universit\"at Innsbruck,
A-6020 Innsbruck, Austria}
\thanks{E.Pazy, acknowledges support through a Kreitman fellowship.
}}

\maketitle

\begin{abstract}

Utilizing the Pauli-blocking mechanism we show that shining circular polarized light on a singly
charged quantum dot induces spin dependent fluorescence. Employing the quantum-jump technique 
we demonstrate that this resonance luminescence, due to a spin dependent 
optical excitation, serves as an excellent read out mechanism for measuring the spin state of 
a single electron confined to a quantum dot.
\end{abstract}

\IEEEpeerreviewmaketitle

\section{Introduction}

\PARstart{S}{emiconductor self}-assembled quantum dots (QDs) form spontaneously during the epitaxial
growth process, with confinement provided in all three dimensions by a high bandgap in the
surrounding material (for example, see \cite{QD}). The strong confinement in such structures leads to a discrete
atom-like density of states which makes these QDs especially attractive for
novel device applications in fields such as quantum computing, optics, and optoelectronics.
It is therefore tempting to try to import successful methods used in quantum optics to the field of QDs.
As we shall show, one should be cautious in considering the atom­-QD analogy. Unlike electrons in an isolated atom,
self-assembled QDs are embedded in an underlying lattice ,i.e., a three-dimensional crystal structure. Even though the 
number of atoms in such dots is 
small compared to lithographically defined QDs, still this underlying
lattice in which the QD is formed strongly affects the single particle states through
its band structure. 

Due to their discrete density of states, QDs have been suggested as the major building
blocks for numerous quantum computing implementation schemes. These implementation schemes
can roughly be divided into those which utilize the charge \cite{Charge},
or respectively the spin \cite{Spin} of charge carriers confined within a QD. 
Accurate measurement of a single qubit is an essential requirement for the implementation of
quantum computation (QC), therefore highly precise methods for the measurement
of the spin of a QD confined charge carrier need to be devised.
Implementation of a highly efficient solid-state measurement scheme designed to measure the
spin or charge of single electron is an extremely difficult task
\cite{Milburn01}. An important requirement is that the measurement
apparatus do not induce decoherence of qubits while no measurement is taking place.
Laser based measurement schemes are ideal in this respect since there
is no dephasing while the laser field is turned off. There have been 
proposals for optical measurements of charge using laser pulses
\cite{STRIP}. Monitoring the fluorescence from a single quantum dot (QD) has
been suggested as a mean to measure single scattering events
within QDs \cite{Hohenester}, as well as a means for final read
out of the spin state for the purpose of quantum computation
\cite{Imamoglu00}. Recently, it has been verified experimentally
that the spin state of an electron residing in a QD can be read
using circularly polarized light \cite{Cortez}. In this
paper we start with the case with no heavy/light 
hole mixing, which has also been recently treated in
 \cite{Shabeav}, and then we proceed to demonstrate 
how it is still possible to devise an optical read
out scheme even in the presence of heavy/light hole mixing.  

Starting with the Hamiltonian for a singly charged QD 
coupled to the laser field, we consider $\sigma^{+}$-polarized laser pulses 
in resonance with the lowest excitation. Under these
conditions we will show how via the Pauli blocking mechanism, experimentally
verified in QDs \cite{Warbuton97}, we are able to obtain spin conditional coupling
of a QD-confined electron to the laser field. Restricting the single QD Hamiltonian to  
a simplified (ideal) three-level scheme and utilizing the quantum jump method we will
show how fluorescence can be employed to measure the spin of a single electron residing in the QD.
The idea behind this read-out technique is the following:
When the polarized laser pulse is switched on, due to Pauli-blocking only one of the spin
states of the confined electron will optically excite a charged exciton state producing  
coherent oscillations between the electron-photon and trion state, these
being disrupted by spontaneous emission of photons.
Therefore, depending on the initial spin state of the electron, in one case
the luminescence will exhibit bright periods whereas for the other spin state it will remain dark,
i.e. no photons will be emitted from the QD. We will proceed to describe
the effects of heavy/light hole mixing on our proposed measurement implementation
scheme, showing that although mixing invalidates the assumption of perfect Pauli-blocking 
it is still possible to use our measurement scheme to read out the spin state of an electron.
We also consider the effects of finite detection efficiency.

\section{Model}

In the quantum jump technique the dynamics of a quantum system is described by 
a non-unitary time evolution defined by an effective non-Hermitian
Hamiltonian. We start by defining the coherent evolution of a singly occupied QD
coupled to a laser field. 

The general Hamiltonian describing the dynamics of a single QD  
interacting with a classical light field can be schematically written as
\cite{QD}
\begin{equation}
\label{eq:Hamiltonian}
H = H_{0} + H_{cl} = (H_{c}+ H_{cc}) + H_{cl}\ ,
\end{equation}
where $H_c$ accounts for the non-interacting confined carriers,
i.e., electrons and holes, $H_{cc}$ describes the Coulomb interaction between
charge carriers and $H_{cl}$ describes the interaction with the classical light field.

\subsection{Single particle states}
To demonstrate the validity of our proposed measurement scheme we
employ the traditional approach to obtaining single 
particle states, utilizing the envelope function approximation \cite{QD,Cardona}
combining it with the ${\bf k \cdot p}$ approximation \cite{Kane}, in which the
QD wave function are defined in terms of $ \Gamma$-like bulk band edge states.
Utilizing the above method we classify the single particle states
according to the value of $(|\mathbf{S }+\mathbf{m}|,S_{z}+m_{z})$,
where $S_{z}$ and $m_{z}$ are projections of the spin and internal band angular momentum
in the crystal growth direction.
We obtain the states $(3/2,\pm 3/2)$  --~heavy-hole (HH) 
subband, the $(3/2,\pm1/2)$ states --~light-hole (LH) subband-and the  $(1/2,\pm 1/2)$ 
--~spin-orbit split-off subband-the latter of which we can safely ignore being energetically far apart.
It is convenient for later calculations of the dipole matrix element
to introduce the four $ \Gamma$ point Bloch functions \cite{Bastard} which serve as a basis
for the crystal states with energies corresponding to the top of the occupied valence band or the bottom
of the conduction band. 
These states are labeled: $\left| S\right\rangle,\left| X\right\rangle,
\left| Y\right\rangle,\left| Z\right\rangle$. The following are the expressions for
the HH and LH states in terms of the Bloch functions and the spin of the functions:

$HH :\left\{ \begin{array}{rlr}
(3/2, 3/2)\leftrightarrow & \frac{1}{\sqrt{2}}\left| (X + \imath Y) \uparrow \right\rangle \\
(3/2, -3/2)\leftrightarrow & \frac{1}{\sqrt{2}}\left| (X - \imath Y)  \downarrow \right\rangle
\end{array} \right. $\\
$LH: \left\{ \begin{array}{rll}
(3/2, 1/2)\leftrightarrow & -\sqrt\frac{2}{3}\left| Z \uparrow \right\rangle+
\frac{1}{\sqrt{6}}\left| (X +\imath Y) \downarrow \right\rangle \\
(3/2, -1/2) \leftrightarrow & -\sqrt\frac{2}{3}\left| Z \downarrow \right\rangle-
\frac{1}{\sqrt{6}}\left| (X -\imath Y) \uparrow \right\rangle
\end{array}
\right. $
\\

Further more we employ the effective mass approximation and 
strain-induced effects are not treated explicitly, rather, we consider an effective confinement potential.

It is important to stress that our proposed measurement scheme does not relay on the above approach for obtaining the single particle states,
we could just as well employed an atomistic psuedo-potential approach\cite{Zunger}. The necessary
requirements which need to be satisfied in order to employ our scheme are that single particle states
are well separated, energetically, and have well defined angular momentum values.

\subsection{Qubit states}
The states we are interested in measuring via the quantum-jump technique are the 
spin states of an excess electron confined to a QD. These states serve as
qubits for several schemes of quantum computation \cite{Spin,Pauli}.
We label the spin states of the excess electron with
\begin{eqnarray}
\left| 0\right\rangle  &\equiv &c_{0,-1/2}^{\dagger
}\left|
\mathrm{vac}\right\rangle, \label{def0}\\
\left| 1\right\rangle  &\equiv &c_{0,1/2}^{\dagger
}\left| \mathrm{vac}\right\rangle ,
\label{def1}
\end{eqnarray}

where $c_{0,\sigma}^{\dagger}(c_{0,\sigma})$ denote creation (annihilation)
operators for electrons in their single-particle ground states with spin projections $\sigma$
and $\left|\mathrm{vac}\right\rangle$ stands for the electron-hole vacuum, i.e., the crystal ground state.
In terms of Bloch functions these states are given by:
$\left| 1\right\rangle \leftrightarrow  \imath \left| S \uparrow \right\rangle ,
\left| 0\right\rangle \leftrightarrow \imath \left| S \downarrow \right\rangle $
These are eigenstates of the bare Hamiltonian $H^c$, with
eigenvalues $\epsilon^e_{0,-1/2}$ and $\epsilon^e_{0,1/2}$
respectively, and are not affected by the carrier-carrier
interaction $H^{cc}$. 

\subsection{Pauli blocking}
  
\begin{figure}
\centering
\includegraphics[width=2.5in]{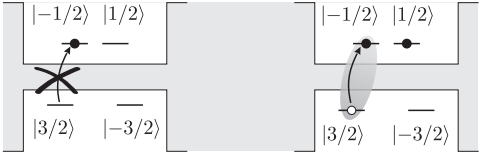}
\caption{Pauli-blocking mechanism: a pulse of
$\sigma_+$-polarized light can promote an electron from the
valence band to the conduction-band $-1/2$-spin state of a quantum
dot only if the latter is not occupied, i.e., if the excess
electron in the dot is in the opposite spin state (right).
Otherwise, no excitation takes place (left).}
\label{fig:1}
\end{figure}

On shining, in the growth direction, a $\sigma^{+}$ polarized classical light laser pulse defined by
an amplitude $E(t)$ and central frequency $\omega _{L}$ on a QD with
an excess electron in the ground state, if the laser is tuned on the lowest interband
excitation energy, a ground-state exciton can be obtained.
The Pauli exclusion principle forbids double
occupancy of any of the electronic states. In particular, if the
excess electron occupies the state $\left| 0\right\rangle $, no
further electron can be promoted from \ the valence
band into that state, and thus creation of a charged exciton by a $\sigma ^{+}$%
-polarized laser pulse is inhibited (left part of Fig.\ref{fig:1}). 
On the other hand, if the excess electron
was in $\left| 1\right\rangle $, \
nothing could prevent a second electron from being excited to the  
spin down state, thereby creating the charged exciton state 
(right part of Fig.\ref{fig:1}).
We consider the absolute value of the Rabi frequency, defined as
\begin{equation}
\Omega (t)\equiv \frac{2\mu _{00}^{-1/2,+3/2}E(t)}{\hbar },
\end{equation}
(where $\mu_{00}$ denotes the dipole matrix element between
the electron and hole ground state wave functions),
to be much smaller than the intraband excitation energy, $\left|
\Omega \right| \ll \omega _{e,h}$. Then we can safely neglect the
probability of promoting the electron from the valence band to a
higher-excited conduction band state. Under these assumptions, the
interaction Hamiltonian simplifies to
\begin{equation}
H_{cl}=\hbar \Omega (t)e^{-i\omega _{L}t}c_{0,-1/2}^{\dagger }d_{0,+3/2}^{\dagger }+h.c. ,
\end{equation}
where $d_{0,\sigma}^{\dagger}(d_{0,\sigma})$ denote creation (annihilation)
operators for holes in their single-particle ground states with spin projections $\sigma$.
The spin polarization of the
created electron-hole pair ($-1/2, 3/2$) respectively, is due to the $\sigma^{+}$ polarization of the light field.
If the temperature is sufficiently low with respect to the
electronic intraband excitation energy, $k_{B}T\ll \hbar \omega
_{e}$, then we can neglect also the excited states of the excess electron.

\subsection{Trion state}
Since we will be considering QDs in the strong confinement regime, i.e., 
the QD level spacing is the largest energy scale in the problem, a good approximation is to assume 
that the system remains in its ground state.
The above consideration allows us to consider a single-QD three-level 
sub-space scheme: two states which correspond to the two spin states
of an excess electron in the QD ($ \left| 0\right\rangle, \left| 1\right\rangle$) and the third state 
is the the so-called ``trion'' --~i.e., the state obtained from
Eqs. (\ref{def0}-\ref{def1}) by creating a charged exciton, $ \left| x\right\rangle$.
The effective Hamiltonian in the three level sub-space is:
\begin{equation}
H_{0} =-\Delta \left| x\right\rangle \left\langle x\right| +\left (\frac{\Omega }{%
2}\left| 1\right\rangle \left\langle x\right| +h.c.\right ) \ ,
\end{equation}
where  $\Delta$ is the detuning from the $\left|
1\right\rangle \rightarrow \left| x \right\rangle $ transition.
The detuning also includes a part which is due to Coulomb interactions present in the trion
state. The Coulomb interactions present in the trion state
modify the ``bare'' trion energy
$\epsilon^e_{0,1/2}+\epsilon^e_{0,-1/2}+\epsilon^h_{0,\sigma_h}$.
The interaction $H^{cc}$ changes the bare state $c_{0,+1/2}^{\dagger
}c_{0,-1/2}^{\dagger }h_{0,\sigma _{h}}^{\dagger }\left| \mathrm{%
vac}\right\rangle$ into the physical interacting state 
$\left|x,\sigma _{h}\right\rangle $, where $\sigma_h$ is the
total angular momentum projection of the hole state. Such states were observed and
studied experimentally in single self-assembled QDs \cite{trions}.
This three-level subspace, spin selectively coupled to a laser
field, was also the basis for a quantum implementation scheme which
we previously proposed \cite{Pauli}.
For further details on obtaining the this effective Hamiltonian and their single particle
states, the reader is referred to \cite{Deco}.

\subsection{Coulomb interaction}

With regards to the carrier-carrier interaction part of the Hamiltonian,
Eq. (\ref{eq:Hamiltonian}), it should be noted that, as opposed to higher 
dimensional quantum structures, in QDs carrier-carrier interactions only induce an
energy level renormalization without causing scattering or dephasing. The carrier-carrier interaction
is small, its smallness being expressed by the
parameter $L/a_B$, where $L$ is the typical spatial dimension of the QD and
$a_B $ is the effective Bohr radius.

\subsection{Incoherent part: exciton recombination}
\label{subsec:incoherent}
The trion state contains two electrons and a hole, an electron and
the hole can recombine and emit a photon. The radiative recombination rates 
are calculated by the Fermi Golden Rule under the dipole approximation.
In this approximation the recombination is
controlled by the dipole matrix element defined by:
$ \langle f | {\bf \epsilon \cdot p} | i \rangle$ where $ |i \rangle$ and
$|f \rangle$ are the initial and final states respectively. The possible 
recombination channels are determined by symmetry considerations 
regarding the dipole matrix element. Using the 
Bloch functions to represent the states one can see that
the only non-vanishing transitions are
$\langle S|x|X\rangle, \langle S|y|Y\rangle,\langle S|z|Z\rangle$.
Due to the symmetry of the QD we assume
$\langle S|x|X\rangle \sim \langle S|y|Y\rangle$ but the ratio of these matrix elements to 
$\langle S|z|Z\rangle$ depends on many factors, e.g. QD shape and size. 

From the above considerations we see that HH transitions only occur via emission of polarized light.
Without the mixing of the hole states an, electron-hole pair composing the trion state
can only recombine emitting a $\sigma^{+}$ polarized photon.
When there is mixing of the HH and LH states, there is a further decay channel 
via the emission of a linear polarized photon, $\sigma^{0}$. 
This can be seen by considering the 
following matrix elements:
$\langle \frac{1}{2},\frac{1}{2}|z|\frac{3}{2},\frac{1}{2} \rangle \sim  \langle S|z|Z\rangle , 
\langle \frac{1}{2},-\frac{1}{2}|z|\frac{3}{2},-\frac{1}{2} \rangle  \sim  \langle S|z|Z\rangle$.

\section{State read-out by quantum jumps}

Our proposed measurement scheme is based on shining a continuous
$\sigma^{+}$ polarized laser pulse in resonance with the trion excitation
on a QD containing an electron in the ground state.
Continuously monitoring of the luminescence from the the QD we expect
to see photons emitted from the QD, i.e. fluorescence,
 if the initial spin state of the excess electron in the QD
corresponds to $|1 \rangle$. If instead the excess electron
is initially in the spin state corresponding to $|0 \rangle$,
we expect to see no photons, i.e., dark periods. This measurement scheme
belongs to the category of indirect measurements \cite{Meas} which are
composed of a two-step
process: first the system, corresponding to the confined
electron in the above scheme, is brought in contact with a ``probe'' quantum system 
 prepared in an initial state, which corresponds to the laser field
 prepared with a given polarization.
The second step is the direct measurement of some observable of the probe which
in the above scheme simply means measuring the emitted photon.
Our proposed measurement scheme follows the requirements from a high precision
indirect measurement: First, the direct measurement of the quantum probe,
(the photon) does not begin before the first step is complete,
(the photon carrier interaction is over). Second (in the non-mixing case which is defined
later on), the measurement of the probe observable does not contribute significantly
to the total measurement error, i.e., we have a high ability to detect the 
emitted photon. Having met the above requirements, the only source of error in our 
measurement scheme arises from the internal uncertainties of the
quantum probe, e.g., the exact polarization of the laser pulse.
In the following we will assume that the laser pulse can be shined in a given
direction and with an exact polarization. We shall also assume in the next
sections perfect photon detection, i.e. that every photon emitted from the QD
is detected with probability 1 which we describe by a detection efficiency
$\eta=1$. The latter of course 
is an unrealistic limit and we will discuss in detail what happens when the 
photon detection is less then 1. Using an avalanche photo-diode detector
the detection efficiency is about 80~\% (for example see \cite{Bachor}). 
The typical wave length emitted by the
recombination process in the QDs lies well within the spectral
window which is due to the cutoff by band gap energy of such
detectors. The main source for low detection efficiency is due to
the probability for the emitted photon to reach the detector, i.e.
the difficulty arising due to finite angle coverage of the
detector. The situation however can be significantly improved by
coupling the QD with a microcavity as described in
\cite{Imamoglu00}.

The measurement scheme is theoretically described by the quantum-jump
technique. Employing this technique we write down an effective non-Hermitian
Hamiltonian describing the dynamics of a reduced density matrix $\rho$.
The typical time associated with this measurement scheme
is the time of measurement $T_{M}$ which is the time the laser 
pulse is switched on.
We choose the measurement time to be such that it minimizes the 
measurement error which is a sum of two factors: the  first is the possibility that 
starting with an electron in state $|1 \rangle$ still during the 
whole measurement time no photon was detected from the QD.
We denote the probability for such an error by $Er_{1}(t)$. The second
kind of error, denoted by  $Er_2(t)$, is that starting the system off in state $|0 \rangle$
a photon was emitted (and detected) from the QD. $T_{M}$ is chosen
so that it minimizes the sum of $Er_1(t)+Er_2(t)$. We start by describing the
measurement process in the case with no HH, LH mixing.

\subsection{No mixing scenario} 
Cases with no HH LH mixing have been experimentally observed
in InP QDs \cite{Efrat} as well as in resonant excitation experiments \cite{Paillard,Lenihan} for InAs/GaAS  which showed that carrier spins are totally frozen on exciton life time scale. The case with no
mixing is defined by $\varepsilon=0$, were $\varepsilon$
is the mixing parameter to be defined later. Shining a $\sigma^{+}$ pulse on the
QD we obtain due to the Pauli blocking effect in QDs the usual
two-level situation: no fluorescence from initial state
$|0\rangle$, full fluorescence from state $|1 \rangle$. 
Assuming perfect photon detection $\eta=1$, one 
measurement is a one shot measurement, 
i.e. one has to wait long enough to ensure that if starting with the
qubit in state $\left| 1\right\rangle$ a single photon will be emitted,
but once this photon is emitted the system is known to be in state 
$\left| 1\right\rangle$.
If $\eta<1$ one has to wait long enough to make sure a detected
photon will be observed, that is, if the electron was in state
$\left| 1\right\rangle$ one has to wait long enough to ensure enough
photons will be emitted so as at least one of them will be detected.

The limiting process which determines the errors
$Er_1$ and $Er_2$ described above is the spin
coherence time in the QD.
Therefore in this case the measurement time, $T_{M}$, is controlled by
the typical time in which a spin flip transition occurs 
in the QD. The issue of spin dephasing times in QDs is still
unresolved, the reason being that experimental verification
of theoretical estimates is still dearly needed. Moreover
the spin dephasing time strongly depends on the size and
shape of the QD. Theoretical estimates range from 1-10 ns \cite{Efros}
to smaller than  $\mu$s  \cite{spinde}. 

In the following we shall assume the spin flip rate to be in the $\mu$s range.
A detailed treatment of the recycling based measurement scheme 
for the case of spin relaxation of the order of ns was given by Shabeav et. al. \cite{Shabeav}
who showed that the spin relaxation is suppressed by coupling to light in the strong 
coupling regime, which can be viewed as a sort of quantum Zeno like effect.

The fluorescence pattern is governed by \cite{Zoller86}
\begin{equation}
\label{eq:fluo}
\kappa_{rec} \tilde{\rho}_{x x}^{(0)}(t|t_0)=-{d \over dt}
\left [\tilde{\rho}_{11}^{(0)}(t|t_0)+\tilde{\rho}_{x x}^{(0)}(t|t_0) \right ],
\end{equation}
where $\tilde{\rho}^{(0)}$ is the reduced density matrix for a two level system, $|1 \rangle, |x\rangle$  
in the sub-space where no photons have been emitted since $t_0$ and we estimate the recombination rate
$\kappa_{rec}\approx 10^9 sec^{-1}$. 
Eq (\ref{eq:fluo}) describes the decay of the trace of $\tilde{\rho}$, i.e., the probability to emit a photon.
One continues to obtain photons (or no photo-emission) as long as the 
original spin state of the electron has not been flipped. Considering the typical time for a spin flip to be 
in the $\mu$s range \cite{spinde} the average number of photons emitted in a fluorescence pattern, which is given by the ratio of the
spin coherence time to the typical rate for spontaneous emission, should be of the order of $10^3$.
In this the detection efficiency can be made greater by using detectors which do not need to detect
single-photon events \cite{Bachor}. 

\subsection{Case with mixing} 
Even though in recent experiments in InAs/GaAs QDs
\cite{Paillard,Lenihan} exhibit no decay of both linear and
circular luminscence polarization, indicating that there is 
apparently no HH LH mixing in these QDs,
typically self-assembled QDs will exhibit mixing
of the HH and LH states.
Such mixing invalidates the assumption of perfect Pauli blocking with $\sigma^+$ light and can
be viewed as a rotation of the basis by an angle $-\varepsilon$ in the $\{|0
\rangle, |1 \rangle\}$ space. The mixing parameter $\varepsilon$,
depends on the material and the sample, and can reach values up to the 
order of 10\% which is the value 
we employ for our calculations.

\begin{figure}
\centering
\includegraphics[width=2.5in]{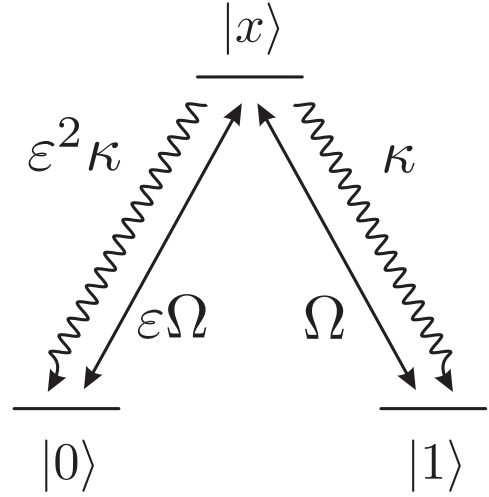}
\caption{The lambda configuration one has to
consider to include hole mixing.}
\label{fig:2}
\end{figure}

Introducing mixing requires one to treat the full three-level
lambda configuration shown in Fig. \ref{fig:2}. As opposed to the
usual atomic lambda configuration \cite{Zoller87}, here one can
not distinguish between the $|0 \rangle \langle x|$ and $|1
\rangle \langle x|$ transitions. These two transitions are
mediated through the same photon. This is basically the signature
of the back action of the photon on the measured spin of
the electron. 

Employing the quantum-jump technique the dissipative evolution of the
density matrix $\tilde\rho(t)$, in the case were the photon is a
$\sigma^+$ light pulse shined in the growth direction,
is given by (see, e.g.,\cite{Quantumnoise})
\begin{eqnarray}
\dot{\tilde\rho}_{00}&=&i\frac{\Omega}2\varepsilon\left(\tilde\rho_{x0}-\tilde\rho_{0x}\right)\nonumber\\
\dot{\tilde\rho}_{11}&=&i\frac{\Omega}2\left(\tilde\rho_{x1}-\tilde\rho_{1x}\right)\nonumber\\
\dot{\tilde\rho}_{xx}&=&i\frac{\Omega}2\left[\tilde\rho_{1x}-\tilde\rho_{x1}+
\varepsilon\left(\tilde\rho_{0x}-\tilde\rho_{x0}\right)\right]-(1+\tilde{\varepsilon}^2)\kappa\tilde\rho_{xx}\nonumber\\
\dot{\tilde\rho}_{01}&=&i\frac{\Omega}2\left(\varepsilon\tilde\rho_{x1}-\tilde\rho_{0x}\right)\label{eqrho}\\
\dot{\tilde\rho}_{0x}&=&i\frac{\Omega}2\left[\varepsilon\left(\tilde\rho_{xx}-\tilde\rho_{00}\right)-\tilde\rho_{01}\right]
-(1+\tilde{\varepsilon}^2)\frac\kappa
2\tilde\rho_{0x}\nonumber\\
\dot{\tilde\rho}_{1x}&=&i\frac{\Omega}2\left(\tilde\rho_{xx}-
\tilde\rho_{11}-\varepsilon\tilde\rho_{10}\right)-(1+\tilde{\varepsilon}^2)\frac\kappa
2\tilde\rho_{1x}\nonumber
\label{eq:densitymat}
\end{eqnarray}
where $\tilde{\varepsilon}^2=\varepsilon^2+\varepsilon'^2$ and $\varepsilon'$ is a result of a further recombination channel,
described below, which is 
allowed due to the mixing. Introducing mixing of the HH and LH states affects both the coherent 
and the incoherent (recombination) part of the Hamiltonian in different ways. 
In the case where there is a single $\sigma^+$ light pulse shined in the growth direction, the ratio
of the Rabi frequencies for the $|1 \rangle \langle x|$ and $|0\rangle \langle x|$ transitions
is given by $\varepsilon$, but the incoherent recombination rates
are not simply given by $\varepsilon^2$. The reason for this is that a further
recombination channel is ``opened up'' due to mixing.
The allowed recombination transitions are restricted by allowed
photon emitted states which should have a total angular momentum of one.  
Therefore there are two possibilities for such a recombination process
(see Sec\ref{subsec:incoherent}). Since the second decay channel is again
proportional to the mixing parameter squared, $\varepsilon^2$, but
with a different coefficient determined by the dipole matrix element
$\langle S|z|Z\rangle$, we describe its effect on the decay by adding
to $\varepsilon^2$ a further $\varepsilon'^2$ term. In a previous paper
\cite{Deco} we neglected this further decay channel here we considered 
both cases: $\varepsilon'=0$ and the opposite limit 
$\varepsilon'\sim \varepsilon$ in which we take $\tilde{\varepsilon}=\sqrt{2}\varepsilon$.

\begin{figure}
\centering
\includegraphics[width=2.5in]{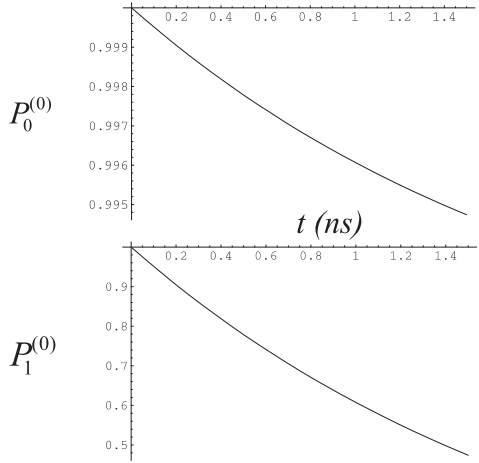}
\caption{Probability that at time $t$ the first
photon has not yet been emitted, starting from state $|0\rangle$
(above) or $|1\rangle$ (below) at time $t=0$. Parameters are
quoted in the text.}
\label{fig:3}
\end{figure}

The probability that at time $t$ no photon has been emitted,
starting from state $\alpha$ at time $t_0$, is
\begin{equation}
P^{(0)}_\alpha(t-t_0)={\rm tr}\left[\tilde\rho(\alpha,t)\right],
\label{eq:prob}
\end{equation}
where at the initial time $t_0$ we take
$\tilde\rho(\alpha,t_0)\equiv|\alpha\rangle\langle\alpha|$.
Fig.~\ref{fig:3} shows an example of their evaluation with
$\Omega=3$ meV, $\kappa=1$ ns$^{-1}$ and $\varepsilon=\tilde{\varepsilon}=0.1$.
For comparison we took the case (same parameters) in which 
${\tilde{\varepsilon}\over \sqrt{2}} =\varepsilon=0.1$. In this case we
got exactly the same results as in Fig.~\ref{fig:3}. As can be seen from Eq.
Eq.~(\ref{eqrho}) the $\tilde{\varepsilon}^2$ term can be neglected
compared to 1. It is also worthwhile to notice that $P^{(0)}_1(t)$ 
does not decay to zero as in the case $\varepsilon=0$ rather both $P^{(0)}_1$ and $P^{(0)}_0$ 
decay to some constant value asymptotically since states $|0\rangle$ and 
$|1\rangle$ no longer correspond to the ``dark'' and ``fluorescent'' states
respectively, i.e. there is also a ``dark side'' to state $|1\rangle$.

\begin{figure}
\centering
\includegraphics[width=2.5in]{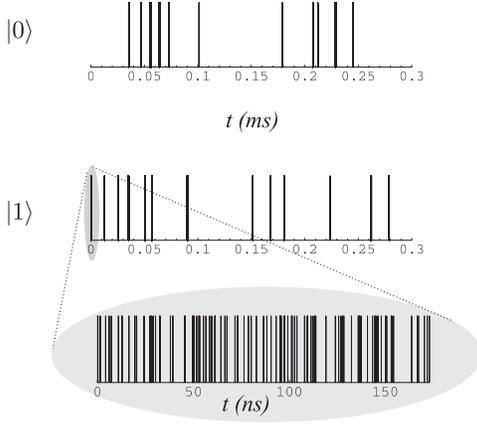}
\caption{Simulation of photon counts for a system
starting from state $|0\rangle$ (top) and from state $|1\rangle$
(middle). An expanded view of the first few photon counts is
displayed in the bottom graph. Parameters are the same as in Fig.
\protect\ref{fig:3}.}
\label{fig:4}
\end{figure}

There are a few major differences in the above case with mixing 
in contrast to the ``common'' lambda configuration
\cite{Zoller87}, commonly used in quantum optics.
In Eq.~(\ref{eqrho}) both of the recycling terms,
$\kappa\tilde{\rho}_{xx}$ and
$\varepsilon^2\kappa\tilde{\rho}_{xx}$, are missing, since it is
the same photon that induces both these transitions, i.e. we can
not distinguish between the two transitions via photon detection.
This implies that, when the first photon is emitted, say at time
$t_1$, the system collapses either into state $|0\rangle$ --~with
probability $p_0=\tilde{\varepsilon}^2/(1+\tilde{\varepsilon}^2)$~-- or into state
$|1\rangle$ --~with probability $p_1=1/(1+\tilde{\varepsilon}^2)$~--,
whence the evolution starts over again. Therefore the probability
that, at the time $t>t_i$ ($i\geq 1$), the $(i+1)$-th photon has
not been emitted, is
\begin{equation}
\label{Pi} P^{(i)}_\alpha(t-t_i)=\frac{\tilde{\varepsilon}^2
P^{(0)}_0(t-t_i)+P^{(0)}_1(t-t_i)}{1+\tilde{\varepsilon}^2},
\end{equation}
which is independent of the initial state $|\alpha\rangle$. 
The possibility for the emitted photon to induce a flip adds to the back-action
of the quantum ``probe'' on the measured system. The measurement time
is again limited by the time the life time the quantum ``memory'' ,i.e.,
by the typical time a spin flip will occur. But in this case
the spin flip will occur with high probability due to repeated photon emissions.
The measurement time is therefore back-action limited.
A typical photo-emission pattern will look like Fig.~\ref{fig:4}: a
sequence of pulses, each one made out of a bunch of the order of
$1/\tilde{\varepsilon}^2$ photons, separated by no-emission windows. This
is the typical quantum-jump pattern one obtains in the presence of
an emission probability having the form of a sum of different
exponentials like Eq.~(\ref{Pi}). 
$N$, the average number of photons in the first bunch, is given 
by a random walk like calculation:
$ N =\sum_{n=0}^{\infty} n q^n p= q {\partial \over \partial q}
\sum_{n=0}^{\infty} q^n p$, where $q=1/(1+\tilde{\varepsilon}^2)$ and
$p=1-q$ stand for the probabilities to decay into states $|1 \rangle$ and
$|0 \rangle$ respectively. Thus one can easily see $N \approx \tilde{\varepsilon}^{-2}$.
In our case as discussed above (see Fig.~\ref{fig:4}) it is of the
order of $\tilde{\varepsilon}^{-2}= 50$. It is only through this first
bunch of photons, which are emitted almost
immediately in the case of state $|1\rangle$, and after a sensible
delay in the case of state $|0\rangle$, that one can discriminate the two patterns.

Starting with the system in
state $|1 \rangle$ there is a possibility for no photon to be
emitted from the QD during the whole measurement time. The
probability for this type of error is given by $Er_1(t)=P^{(0)}_1(t)$. In
the other case starting with the system initially in $|0 \rangle$
at least one photon might be emitted during the measurement time.
The probability for this sort of error is given by:
$Er_2(t)=1-P^{(0)}_0(t)$. The measurement time has to be chosen in such a
way as to minimize the sum of these two errors. For the same
parameters employed in Fig.~\protect\ref{fig:3}, we obtain an
estimate for the optimal measurement time of the order of a few
tens of ns. What typically happens in practice is that, as shown
in Fig.~\ref{fig:4}, by appropriate time windowing the first
bunch of photons coming from state $|1\rangle$ can be safely
discriminated from the (later) photons coming from state
$|0\rangle$. 

\subsection{Two laser pulses}
One can try to improve the detection capability by shining more than 
one laser pulse on the QD.
Shining a further laser pulse, phase matched to the first, in the in-plane direction with a $\hat{z}$ 
linear polarization, $\sigma^{0}$, 
couples only to the LH wave function since 
only they posses a $|Z \rangle$ Bloch function component.  There are two Rabi frequencies in this case: $\Omega$ for the 
$\sigma ^{+}$ polarized light shined in the growth direction $\hat{z}$ and  $\Omega'$ for the linear polarized laser
pulse (in the $\hat{z}$ direction) shined in the in-plane direction, e.g. $\hat{x}$. The modified equations of
motion can be obtained from Eq(\ref{eq:densitymat}) by modifying the coherent part of the Hamiltonian,
i.e., the Rabi frequencies. The modified Rabi frequencies
are: $\Omega \rightarrow \Omega \pm  \varepsilon \Omega' $ 
and $\varepsilon \Omega \rightarrow \varepsilon (\Omega \pm \Omega')$. 
This can be seen by considering the following matrix elements:
$\langle \frac{1}{2},\frac{1}{2}|z|\frac{3}{2},\frac{1}{2} \rangle \sim  \langle S|z|Z\rangle , 
\langle \frac{1}{2},-\frac{1}{2}|z|\frac{3}{2},-\frac{1}{2} \rangle  \sim  \langle S|z|Z\rangle$,
which show that the LH part of the mixed HH LH state will couple to the conduction band state.
The Rabi frequency for the coupling being $\Omega' $ and $\varepsilon$ is the LH ``part'' of
the mixed state which is coupled. 

Since the further laser pulse modifies only the coherent part of the Hamiltonian 
it will affect the measurement efficiency through the change of the time of measurement
and the value of $Er_2(T_{M})$. The bottleneck
process limiting the measurement time $T_{M}$ is the finite probability
$p_0=\tilde{\varepsilon}^2/(1+\tilde{\varepsilon}^2)$, for decay into 
state $|0\rangle$ every time a photon is emitted. This finite probability
is controlled by the decay rate, i.e., the incoherent part of the Hamiltonian,
and as such will not be affected by the second lase pulse. Thus the first
bunch of photons emitted from the initial state $|1\rangle$ will be
of the order of $1/\tilde{\varepsilon}^2$ photons, independent on the 
number of laser pulses. The measurement time, $T_{M}$ is affected by the coherent
part of the Hamiltonian since it determines how quickly a photon will be emitted
when the system starts off in state $|1\rangle$. The error rate $Er_2(T_{M})$ is
also affected by the coherent part. Therefore one can see the optimal case
will be achieved for $\Omega=-\Omega'$.

\subsection{Finite detection efficiency} We now consider the case
in which $\eta <1$. For this case there is a further measurement
error denoted by $Er_3(t)$. This error is due to the possibility
that starting off initially in
state $|1 \rangle $ the QD can emit a photon/photons which will go
undetected and the spin can flip into state $|0 \rangle$, i.e.,
the information regarding the spin state is lost without being
detected.
The probability for such an error is given by
\begin{eqnarray}
Er_3&=& \tilde{\varepsilon}^2\sum_{n=0}^{N} \left(\frac{1-
\eta}{1+\tilde{\varepsilon}^2}
\right)^{n+1}\nonumber\\
&=&\frac{\tilde{\varepsilon}^2 (1- \eta)}{\tilde{\varepsilon}^2 +\eta}\left [1-
{\left ({1- \eta \over 1+\tilde{\varepsilon}^2}\right )}^{N+1}\right ],
\end{eqnarray}
which is simply the sum over $n$ incidents in which the emitted
photons were not detected and no spin-flip occurred and on the
$n+1$ incident such a spin flip occurred (without the photon being
detected). 
Taking a relatively low detection  efficiency $\eta=0.8$ we obtain an error due
to finite detection efficiency less than 0.5~\%.

Working with a detector with a finite efficiency means that we
have to choose the measurement time so as to ensure the
fluorescent state emits a few photons thus increasing the
probability one of them will be detected. This will also increase the
probability for an error due to a photon being emitted by the
initial state $|0 \rangle$ since $P^{(0)}_0(t)$ decays
exponentially with time, but the main error source are
spin flips after undetected photons.

\section{Conclusion}
Employing a quantum-jump technique, we have theoretically demonstrated that it
is possible, utilizing the resonance-luminescence technique, which has been proven to be very effective 
in the field of quantum optics, to measure the spin of an excess electron confined to a QD.
The complications arising from the underlying crystal have been critically examined and we have shown that 
even in the presence of heavy and light hole mixing one is still able to measure the electron's
spin to a very high degree of accuracy.

\end{document}